\documentclass[12pt]{article}
\usepackage{hyperref}
\usepackage{cite}
\usepackage{color}
\usepackage{graphicx}
\usepackage{amsmath}
\usepackage{amssymb}
\usepackage{xspace}

\makeatletter
\@addtoreset{equation}{section}

\makeatletter
\renewcommand\section{\@startsection {section}{1}{\z@}%
                                   {-3.5ex \@plus -1ex \@minus -.2ex}%nn
                                   {2.3ex \@plus.2ex}%
                                   {\normalfont\large\bfseries}}
\renewcommand\subsection{\@startsection{subsection}{2}{\z@}%
                                     {-3.25ex\@plus -1ex \@minus -.2ex}%
                                     {1.5ex \@plus .2ex}%
                                     {\normalfont\bfseries}}

\def\baselinestretch{1.2}
\newcommand{\be}{\begin{equation}}
\newcommand{\ee}{\end{equation}}
\newcommand{\beq}{\begin{eqnarray}}
\newcommand{\eeq}{\end{eqnarray}}

\newcommand{\tr}{{\rm Tr}}
\newcommand{\gone}[1]{{}}

\begin{document}
\begin{titlepage}
\begin{flushright}
%preprint number
\end{flushright}
%\vspace{12 mm}

\vfil
%vfil

\begin{center}

{\Large{\bf Toward Higher Spin ${\bf dS_3/CFT_2}$}}

\vfil

Peter Ouyang

\vfil
{\it Department of Physics, Purdue University\\
525 Northwestern Ave, West Lafayette, IN 47906, USA}
\vfil

\end{center}

%%%%%%%%%%%%%%%%%%%%%%%%%%%%%%%%%%%%%%%%%%%%%%%%%%%%%%%%%%%%%%%%%%%%%%%%%%%%%%%%%%%%%%%
\begin{abstract}
\noindent  I take steps toward the construction of a CFT dual to Vasiliev's higher spin gravity in three dimensional de Sitter space.  There are two main claims.  The first is that higher spin de Sitter symmetries are related to extended Virasoro symmetries, as in AdS; this is verified explicitly for the case of $\mathcal{W}_3$ asymptotic symmetry.  The associated chiral algebra has imaginary central charge.  The second (conjectural) claim, inspired by work of Gaberdiel and Gopakumar in ${\rm AdS_3/CFT_2}$, is that an appropriate CFT can be identified as an exotic non-unitary WZW coset model at complex level.
%I perform a variety of silly calculations.
 
\end{abstract}
%%%%%%%%%%%%%%%%%%%%%%%%%%%%%%%%%%%%%%%%%%%%%%%%%%%%%%%%%%%%%%%%%%%%%%%%%%%%%%%%%%%%%%%%%
\vspace{0.5in}

\end{titlepage}
\renewcommand{\baselinestretch}{1.05}  %Line spacing
%%%%%%%%%%%%%%%%%%%%%%%%%%%%%%%%%%%%%%%%%%%%%%%%%%%%%%%%%%%%%%%%%%%%%%%%%%%%%%%%%%%%%%%%%%%%%

\section{Introduction}

The study of quantum gravity is in a remarkable state.  It was not that long ago that there were no known self-consistent theories of quantum gravity, but happily this is no longer the case.  In flat spacetime, superstring theory gives perturbatively consistent quantum gravity theories with many vacua, while in asymptotically anti-de Sitter spacetime, the AdS/CFT conjecture \cite{Maldacena:1997re,Gubser:1998bc,Witten:1998qj} gives, in principle, a large class of non-perturbatively consistent theories of quantum gravity.

One of the most striking features of AdS/CFT is that the only essential assumption it makes about the gravitational background is that the geometry is asymptotically AdS; the geometry can fluctuate in generic ways in the interior of the space, and in principle the bulk could even have a non-geometric description.  This is a mild assumption for many questions of interest. In particular, AdS asymptotics allow black hole solutions, and their associated singularities, and a large amount of work has gone into studying their properties in light of AdS/CFT.  However, other situations of physical interest are excluded by the assumption of AdS asymptotics.  For example, cosmological solutions are generally not compatible with AdS asymptotics, and one would like in particular to have a theory of quantum gravity capable of describing Big Bang singularities.

Partly for this reason, an analogous gauge theory/gravity duality for de Sitter space has been proposed \cite{Strominger:2001pn,Witten:2001kn,Maldacena:2002vr}.  Unfortunately, it has been difficult to study this idea because of the absence of explicit candidate dualities.  In AdS/CFT, most of the explicitly known duals were constructed with the aid of supersymmetry.  The global symmetries of the theories on each side of the duality must be the same, so with a suitably large amount of symmetry one can often guess the candidate theories on each side of the correspondence and then check that the duality makes sense (indeed, if one were to guess the CFT dual to maximally supersymmetric supergravity in ${\rm AdS_5}$, the only candidate is maximally supersymmetric Yang-Mills theory.)  De Sitter space, however, does not arise as a supersymmetric solution of any known gravitational theory.

Therefore it is natural to consider alternatives to supersymmetry as guides in the search for candidate dS/CFT duals.  The most powerful such symmetries are {\it higher-spin} symmetries (for a recent review on how to avoid no-go theorems for extended spacetime symmetries, see \cite{Bekaert:2010hw}.)  Field theory duals of these higher-spin theories have already been proposed in the context of AdS/CFT.  One such duality is the conjecture of Klebanov and Polyakov \cite{Klebanov:2002ja}, that Vasiliev's higher spin theory in ${\rm AdS_4}$ \cite{Vasiliev:1990en} is dual to the large-$N$ limit of the $O(N)$ vector model in three dimensions (important related work includes \cite{Sezgin:2002rt,Giombi:2009wh,Giombi:2011ya}.)  In a further development, Gaberdiel and Gopakumar \cite{Gaberdiel:2010pz} have proposed a duality in ${\rm AdS_3/CFT_2}$, relating Vasiliev's theory in three dimensions \cite{Vasiliev:1995dn} to a WZW coset CFT.

Recently, in a fascinating paper, Anninos, Hartman, and Strominger \cite{Anninos:2011ui} have proposed an explicit example of a three-dimensional CFT that is putatively dual to a gravitational theory in four-dimensional de Sitter space. 
Specifically, they argued that for Vasiliev's theory in $dS_4$, the dual theory is in fact a Euclidean $Sp(N)$ vector model.  The correlation functions of the CFT are related to those of the $O(N)$ vector model by mapping $N$ to $-N$; in particular this reverses the sign of the central charge.  The CFT is not unitary, but this is not a serious obstacle; it describes the physics of the boundary in the infinite future (or past) and therefore does not inherit a notion of time evolution from the bulk theory (unlike AdS/CFT.)  This conjecture opens a new line of attack on the problem of de Sitter quantum gravity.

The purpose of this brief note is to suggest an analogous duality for ${\rm dS_3/CFT_2}$, by analyzing the symmetries of higher spin gravity in ${\rm dS_3}$ and generalizing the WZW coset construction of Gaberdiel and Gopakumar.  In Section \ref{sec2} we study the asymptotic symmetries of higher spin theories in de Sitter space.  One's natural expectation is that a spin-$N$ higher-spin gravity (that is, containing one field each of spin $s= 2,\, 3,\, 4, \ldots N$) in ${\rm dS_3}$ with a natural set of boundary conditions has as its asymptotic symmetry group one complexified copy of $\mathcal{W}_N$, and central charge
\beq
c= \frac{3i\ell}{2G}
\eeq
where $\ell$ is the de Sitter radius and $G$ is Newton's constant.  Notice that the central charge is (crucially) imaginary\footnote{One way to motivate that the central charge should be imaginary is to recall that one can often map quantities from anti-de Sitter to de Sitter by making the identification $\ell_{AdS}\rightarrow i\ell_{dS}$.  See, for example, \cite{Maldacena:2002vr}.  Also it was pointed out  in \cite{Balasubramanian:2002zh} that an imaginary central charge may be natural for a CFT dual to ${\rm dS_3}$, based on the likely Hermiticity properties of the CFT.}.
We will verify this explicitly for the case of $\mathcal{W}_3$; the structure of the derivation makes it clear that the form of the answer is the same for any $N$.  The analysis presented here essentially follows earlier work \cite{Campoleoni:2010zq,Henneaux:2010xg,Gaberdiel:2011wb,Campoleoni:2011hg} with various factors and signs inserted strategically.  In Section \ref{sec3} we will review the essential features of the work of Gaberdiel and Gopakumar \cite{Gaberdiel:2010pz}, and in Section \ref{sec4} we present a conjecture for ${\rm dS_3}$ and subject it to some very modest consistency checks.  The candidate bulk theory is Vasiliev's theory in three dimensions.  On the field theory side, the relevant CFT is the WZW coset
\beq
\frac{sl(N)_k \oplus sl(N)_1}{sl(N)_{k+1}}\nonumber
\eeq
where the level parameter $k$ is complex,
\beq
k= -N + \frac{i}{\gamma}
\nonumber
\eeq
and one must take the dual limit $N,\gamma \rightarrow \infty$.  Many issues remain to be investigated.  We will conclude by describing some of them.

\section{Asymptotic Symmetries of Higher Spins in $dS_3$}
\label{sec2}

In any physical system, perhaps the most basic question one can ask is what its symmetries are and what charges characterize its states.  In gauge theories such as Yang-Mills theory and gravity, this can be a subtle issue because one must distinguish between symmetries which are true gauge symmetries (and map a physical state to itself) and those which are global symmetries (which map a physical state to a different physical state.)  The technique that one uses to disambiguate between gauge and global symmetries is to compute the asymptotic symmetry group (ASG).  In this section we describe the ASG for a theory in 3 dimensional de Sitter space containing gravity and a spin-3 field.

To set the stage, let us recall the situation for general relativity.  For a given gravitational solution, the global symmetries are the gauge transformations which leave the metric invariant, or in other words they are generated by vector fields which satisfy Killing's equation $\nabla_{\mu} \xi_{\nu} +\nabla_{\nu} \xi_{\mu}=0$.  If one tries to compute the conserved currents associated with these symmetries using Noether's procedure, one finds that the currents simply vanish on shell -- this is the well-known result that general relativity does not have a local stress tensor.  However, in applying Noether's procedure, the variation of the action gives rise to a boundary term, and this boundary term can indeed be associated to a conserved charge \cite{Brown:1992br}.

Therefore, to define the charges of a gravitational system, we should impose boundary conditions on the metric and fields so that these boundary charges are well-defined (in particular, they must not diverge.)  The gauge transformations that are consistent with these boundary conditions fall into two classes.  Either they give nonzero boundary charges, in which case we think of them as asymptotic global symmetries, or they fall off too rapidly and give vanishing boundary charges, in which case we think of them as true gauge transformations.  In general, the asymptotic global symmetries of a given background do not have to be equal to the Killing vectors.
For example, the classic computation of Brown and Henneaux \cite{Brown:1986nw} showed that in ${\rm AdS_3}$, the ASG is enlarged from the $SL(2,R)\times SL(2,R)$ global symmetries of ${\rm AdS_3}$ to be the tensor product of two copies of the Virasoro algebra.  It is worth remembering that the specific asymptotics one uses are explicitly a choice; the bulk theory on its own does not determine the boundary conditions.  In practice, the challenge is to find boundary conditions which eliminate physically uninteresting solutions but are still general enough to contain a broad class of interesting solutions.

We will now compute the ASG and its associated central charge for the $SL(3,C)$ Chern-Simons theory that describes spin-3 higher spin gravity in de Sitter space.  The discussion closely follows \cite{Campoleoni:2010zq,Henneaux:2010xg,Gaberdiel:2011wb,Campoleoni:2011hg}; my only original contributions are to insert factors of $i$ and $-1$ in appropriate places, shuffle indices, and to clarify the boundary conditions somewhat in Section \ref{sec23}.  The explicit analysis presented here is special to the case of spin-3 gravity, but we will argue in Section \ref{sec25} that the results extend to general higher spins, in particular the form of the central charge.

\subsection{Chern-Simons Formulation of $dS_3$ Gravity}

For our purposes it is very convenient to use the Chern-Simons formulation of gravity in three dimensions \cite{Witten:1988hc}.  For de Sitter space, this consists of combining the vielbeins and connection 1-forms into vector fields as
\beq
A &=& \left(\omega_{\mu}^{\;\; a} + \frac{i}{\ell} e_{\mu}^{\;\; a} \right) T_a dx^{\mu},\\
\tilde{A} &=& \left(\omega_{\mu}^{\;\; a} - \frac{i}{\ell} e_{\mu}^{\;\; a} \right) T_a dx^{\mu} .
\label{adef}
\eeq
The $T_a$ are real and satisfy the $SL(2)$ algebra, $[T_a,T_b] = \epsilon_{abc}T^c$.  The indices are raised and lowered with the orthonormal frame metric $\eta^{ab} = {\rm diag}(-1,+1,+1)$ and the connection 1-forms with one frame index $\omega^a$ are related to the usual spin connection by  $\omega_a = \frac12 \epsilon_{abc} \omega^{bc}$.

The relevant action is
\beq
S = \frac{\kappa}{4\pi} \int \tr \left( A\wedge dA + \frac23 A\wedge A \wedge A -\tilde{A}\wedge d\tilde{A} -\frac23 \tilde{A}\wedge \tilde{A}\wedge \tilde{A}\right).
\label{csaction}
\eeq
To relate this action to the Einstein-Hilbert action, we must identify
\beq
\kappa = \frac{i\ell}{4G}
\label{kappa}
\eeq
and one can show that the equations of motion $F=\tilde{F}= 0$ are equivalent to Einstein's equations in empty de Sitter space.
Because the vector fields in (\ref{adef}) are complex, the gauge symmetry is $SL(2,C)$, with only one copy as the two vectors are related by complex conjugation $A = \tilde{A}^*$.  Most of the time it suffices to perform the analysis just for $A$.  The gauge fields and the gauge group are both complex, and so the Chern-Simons action of $A$ is also explicitly complex.  One might ordinarily worry that this complex action would give rise to ghost states in the bulk.  However, this is not the case; the full action including both $A$ and its complex conjugate $\tilde{A}$ is explicitly real\footnote{The structure of the CS action in de Sitter might be related to the proposal of Maldacena \cite{Maldacena:2002vr} that in dS/CFT the CFT partition function computes the wavefunction of the universe in de Sitter.  Physical quantities in dS are then computed by squaring the wavefunction and integrating over final states.  In the CS language, the bulk action is naturally squared, suggesting that one should think of the CFT partition function as the dual of just the CS theory of $A$.  Squaring the wavefunction then amounts to integrating over both $A$ and $\tilde{A}$.}.

The gauge transformations of $A$ are
\beq
A \rightarrow A+ d\xi + [A,\xi].
\eeq
In solving the CS equations, one needs to impose boundary conditions so that the variation of the action is well-defined.
In the following, we make the choice
\beq
A_{\bar{z}}|_{bdry} = 0
\label{bc}
\eeq

Let us translate the de Sitter geometry into the CS language.  The dS metric may be written in the flat slicing as
\beq
\frac{ds^2}{\ell^2} = -dt^2 + e^{2t} dzd\bar{z}
\eeq
where $-\infty < t < \infty$ and the boundary of the space is in the infinite future at $t = + \infty$.  Now, we make the gauge choice:
\beq
A_t = i T_0.
\label{Atgauge}
\eeq
Then for the de Sitter metric, the CS gauge fields are
\beq
A &=& iT_0 dt + e^t (T_2 + i T_1) dz,\\
\tilde{A} &=& - iT_0 dt + e^t(T_2 - i T_1) d\bar{z}.
\label{ApuredS}
\eeq

\subsection{Spin-3 Chern-Simons}

From the Chern-Simons point of view, it is simple to construct extended gravity theories by enlarging the gauge group from $SL(2,C)$ to a different Lie group containing $SL(2,C)$ as a subgroup.  The simplest choice is to take the gauge group to be $SL(N,C)$, in which case the theory contains a tower of spins, $s=2,\,3,\ldots N$.  For simplicity and clarity, we will take $N=3$ but one can also study the case of general $N$.

The $SL(3)$ algebra adds five generators $T_{ab}$ (symmetric and traceless in $a,b$) to the three $T_a$ of $SL(2)$.  Explicitly, the algebra is
\beq
&& [T_a,T_b] = \epsilon_{abc} T^c \\
&& [T_a,T_{bc}] = \epsilon^d_{\;\;a(b} T_{c)d} \\
&& [T_{ab},T_{cd}] = \sigma \left(\eta_{a(c} \epsilon_{d)be}+\eta_{b(c} \epsilon_{d)ae}\right)T^e.
\eeq
The form of the algebra is determined by the Jacobi identity and the symmetry properties of the indices up to a normalization $\sigma$.  For dS theories, $\sigma$ is not so important, because the gauge group is complexified, but we will keep track of it anyway to ease comparison with the literature.
Then one considers a CS theory with gauge field given by
\beq
A = \left(\omega_{\mu}^{\;\; a} + \frac{i}{\ell} e_{\mu}^{\;\; a} \right) T_a dx^{\mu}+
\left(\omega_{\mu}^{\;\; ab} + \frac{i}{\ell} e_{\mu}^{\;\; ab} \right) T_{ab} dx^{\mu}.
\eeq
The  new fields $e_{\mu}^{\;\; ab}$ can be contracted into two vielbeins to give a 3-index object, so it is clear that they correspond to a spin-3 field.  It can be shown that they obey linearized equations equivalent to the free higher spin equations of Fronsdal \cite{Fronsdal:1978rb}.

Obviously any ordinary gravity solution can be embedded in the higher spin theory by setting all the higher spin fields to zero.
In what follows, it is convenient to assemble the Lie algebra generators as
\beq
L_0 &=& iT_0\\
L_{\pm 1} &=& T_2 \pm i T_1\\
W_{\pm 2} &=& T_{22}-T_{11} \pm 2i T_{12}\\
W_{\pm 1} &=& \pm T_{01}- i T_{02}\\
W_0 &=& -T_{00}
\eeq
which obey the standard algebra
\beq
&&[L_m,L_n] = (m-n) L_{m+n}\\
&&[L_m,W_p] = (2m-p) W_{m+p} \\
&&[W_p,W_q] = \frac{\sigma}{3}(p-q) (2p^2 + 2q^2 -pq-8) L_{p+q}.
\eeq
Note that the $L$ and $W$ operators are not naturally real, in contrast with the case in AdS; in particular
$ (L_0)^* = -L_0$.  This fact may be related to the nonstandard Hermiticity conjectured in \cite{Balasubramanian:2002zh}.
%Note that if the $T^a$ are real $3\times 3$ matrices, 
%the $L$ and $W$ operators are not real; in particular 
%$ (L_0)^* = -L_0$.  This fact is perhaps related to the 
%non-unitarity of the putative CFT duals to de Sitter.

\subsection{Asymptotic Conditions}
\label{sec23}

In asymptotically de Sitter space, the analogues of the Brown-Henneaux boundary conditions \cite{Brown:1986nw} for the metric at future infinity are 
\beq
&& g_{z\bar{z}} = \frac{1}{2} e^{2t} + O(1)\label{gzzbar}\\
&& g_{tt} = -1 + O(e^{-2t})\label{gtt}\\
&& g_{zz} = O(1)\label{gzz}\\
&& g_{zt} = O(e^{-2t}).\label{gzt}
\eeq
With these boundary conditions, the asymptotic symmetry algebra is the Virasoro algebra \cite{Strominger:2001pn} (note that a small typo in \cite{Strominger:2001pn} is corrected in \cite{Spradlin:2001pw}.)  We would like to translate these boundary conditions to the Chern-Simons formalism and generalize them to higher spins.

For the spin-2 part of the gauge field, the asymptotics are most efficiently summarized as
\beq
A_t^{(s=2)} &=& L_0 + O(e^{-2t})\label{s2asympt}\\
A_z^{(s=2)} &=&  e^t L_1  + O(e^{-t}) \\
A_{\bar{z}}^{(s=2)} &=& O(e^{-2t}) \\
\left(dA + A\wedge A \right)|^{s=2} &=& O(e^{-2t})
\eeq
with similar expressions for $\tilde{A}$.  It can be shown that these conditions are equivalent to (\ref{gzzbar}-\ref{gzt}). 

For the spin-3 components of $A$, the correct conditions appear to be
\beq
A_t^{(s=3)} &=&  O(e^{-3t})\\
A_z^{(s=3)} &=&  O(e^{-2t}) \\
A_{\bar{z}}^{(s=3)} &=& O(e^{-3t}) \\
\left(dA + A\wedge A \right)|^{s=3} &=& O(e^{-3t}).\label{s3asympt}
\eeq
Note that we only require the CS equations of motion to be satisfied asymptotically; this means that the asymptotic conditions are compatible with the addition of matter fields, provided that the matter density becomes sufficiently dilute at future infinity.

The generic $A$ satisfying the asymptotic conditions (\ref{s2asympt}-\ref{s3asympt}) can be written as
\beq
A &=& e^t L_1 dz + L_0 dt + \frac{2\pi}{\kappa} e^{-t}\mathcal{L}(z) L_{-1} dz + \frac{\pi}{2\kappa \sigma} e^{-2t}\mathcal{W}(z) W_{-2}dz
\label{aform}
\eeq
up to spin-2 components which enter at order $O(e^{-2t})$ and spin-3 components which enter at order $O(e^{-3t})$.  These additional terms fall off at future infinity too rapidly to give rise to nontrivial charges, and we will ignore them.  The normalizations of $\mathcal{L}$ and $\mathcal{W}$ are the same as in \cite{Campoleoni:2010zq} and are chosen so that the generators of the ASG are canonically normalized.

Finally, although we will not consider further higher spins explicitly in this paper, the natural asymptotics for the spin-$n$ components are
\beq
A_t^{(s=n)} &=&  O(e^{-nt})\\
A_z^{(s=n)} &=&  O(e^{-(n-1)t}) \\
A_{\bar{z}}^{(s=n)} &=& O(e^{-nt}) \\
\left(dA + A\wedge A \right)|^{s=n} &=& O(e^{-nt}).\label{snasympt}
\eeq
It is not hard to see that the $SL(N,C)$ Chern-Simons theory with these asymptotics has $N-1$ holomorphic functions' worth of asymptotic solutions, which makes it natural that the general asymptotic symmetry should be $\mathcal{W}_N$.

\subsection{Asymptotic Symmetry Algebra}
\label{sec24}

For the Chern-Simons formulation of gravity, the analogues of the Killing vectors are the gauge transformations $\xi$ which leave the gauge field invariant, $d\xi + [A,\xi] =0$.  However, for the purpose of computing the asymptotic symmetries, we should only require that $\xi$ is such that gauge transformations preserve the form (\ref{aform}), or in other words, given the asymptotic conditions (\ref{s2asympt}-\ref{s3asympt}), we require
\beq
\left(d\xi + [A,\xi]\right)|^{s=2} &=& \frac{2\pi}{\kappa}e^{-t}L_{-1} \,\delta \mathcal{L}(z) dz + O(e^{-2t})\\
\left(d\xi + [A,\xi]\right)|^{s=3} &=& \frac{\pi}{2\kappa\sigma}e^{-2t}\,W_{-2} \delta \mathcal{W}(z) dz + O(e^{-3t}).
\eeq

The gauge transformations which do this may be written as
\beq
\xi = \sum\limits_{m=-1}^{1} \epsilon_m(z) L_m e^{mt} +\sum\limits_{p=-2}^{2} \chi_p(z) W_p e^{pt} 
\label{xiform}
\eeq
provided that the functions $\epsilon_m$ and $\chi_p$ satisfy the relations (to reduce index clutter it is convenient to define $\epsilon_1 \equiv \epsilon, \chi_2 \equiv \chi$)
\beq
\epsilon_0 &=& -\epsilon'\\
\epsilon_{-1} &=& \frac{2\pi}{\kappa}\mathcal{L}\epsilon + \frac12 \epsilon''+\frac{4\pi}{\kappa} \mathcal{W}\chi\\
\chi_1 &=& -\chi'\\
\chi_0 &=& \frac{4\pi}{\kappa} \mathcal{L} \chi +\frac12 \chi''\\
\chi_{-1} &=& -\frac{10\pi}{3\kappa}\mathcal{L} \chi' -\frac{4\pi}{3\kappa}\mathcal{L}'\chi -\frac16 \chi''' \\
\chi_{-2} &=& \frac{\pi}{2\kappa\sigma} \mathcal{W} \epsilon + \left(\frac{2\pi}{\kappa}\right)^2\mathcal{L}^2 \chi +\frac{2\pi}{\kappa}\left(\frac23 \mathcal{L}\chi'' + \frac{7}{12} \mathcal{L}'\chi' +\frac16 \mathcal{L}''\chi\right) +\frac{1}{24} \chi''''.
\eeq
Of course, there can be additional terms in (\ref{xiform}) proportional to the $L_m$ of order $O(e^{-2t})$ and terms proportional to the $W_p$ of order $O(e^{-3t})$, but they give rise to trivial charges.

When all of the above conditions are satisfied the functions $\cal{L}$ and $\cal{W}$ vary as
\beq
\delta_{\epsilon} \mathcal{L} &=& 2\mathcal{L} \epsilon' + \mathcal{L}'\epsilon + \frac12 \epsilon'''
\label{w3eq1}\\
\delta_{\chi} \mathcal{L} &=& 3 \mathcal{W} \chi' + 2 \mathcal{W}'\chi
\label{w3eq2}\\
\delta_{\epsilon} \mathcal{W} &=& 3 \mathcal{W}\epsilon'+ \mathcal{W}'\epsilon 
\label{w3eq3}\\
\delta_{\chi} \mathcal{W}&=& \sigma\left(\frac23 \mathcal{L}'''\chi + 3\mathcal{L}''\chi' +5 \mathcal{L}'\chi'' + \frac{10}{3} \mathcal{L} \chi''' \right)\cr
&& \qquad + \frac{64\pi \sigma}{3\kappa} \left( \mathcal{L}^2 \chi' + \mathcal{L}\mathcal{L}'\chi\right) +\frac{\kappa\sigma}{12\pi} \chi'''''\label{w3eq4}.
\eeq
These are the defining equations of the $\mathcal{W}_3$ algebra, with central charge $c =6\kappa$.

The charges associated with these symmetry transformations are of the form
\beq
Q(\xi) = -\frac{\kappa}{2\pi} \oint dz \tr \left(\xi A_z\right)|_{t\rightarrow \infty}.
\label{defQ}
\eeq
They generate the symmetries through
\beq
\delta_{\xi} F = \left\{ Q(\xi),F\right\}
\eeq
where the braces $\{\;,\;\}$ represent Poisson brackets.  By representing the symmetry transformations in (\ref{w3eq1})-(\ref{w3eq4}) as Poisson brackets and Laurent expanding the result, one obtains the $\mathcal{W}_3$ algebra in terms of the more familiar mode expansion.  

There are three comments in order about the charges defined in (\ref{defQ}).  The first is that they are not really ``conserved'' charges in the usual sense, as they are only properly defined on the boundary at future infinity, but rather they are characteristic data labelling the final state of the universe.  The $Q(\xi)$ are only conserved in the sense that they vary exponentially slowly for large but finite $t$. 
The second, related comment is that one can construct the charges (\ref{defQ}) by following a Noether-type argument, but in doing so one obtains an additional term of the form $\int \tr (\xi A\wedge A)$.  The boundary conditions (\ref{s2asympt}-\ref{s3asympt}) guarantee that these extra contributions vanish when the integrals are evaluated at future infinity.  The third comment is that to define the integral $\int dz$ properly, one must choose an integration contour on the future boundary.  In ${\rm AdS_3}$ there is a more-or-less natural contour along the boundary at fixed time, but in ${\rm dS_3}$ any closed contour suffices.  One might worry that the ability to pick any contour gives a degeneracy of charges, but this is not the case -- for a given contour, the various choices of $\epsilon$ and $\chi$ pick off different combinations of the Laurent coefficients of $\mathcal{L}$ and $\mathcal{W}$. So the charges are really just these Laurent coefficients and the charges defined through different contours must be equivalent.

In de Sitter space, the Chern-Simons level $\kappa$ is imaginary (\ref{kappa}) and so is the central charge which one associates with the asymptotic symmetry algebra:
\beq
c = \frac{3i\ell}{2G}
\eeq
The normalization of the central charge is fixed by the form of the Virasoro algebra, up to an overall sign ambiguity (perhaps the simplest way to think about the sign ambiguity is to make a field redefinition interchanging $A$ and $\tilde{A}$, while appropriately exchanging $z$ and $\bar{z}$.)  In particular, we might have tried to obtain a real central charge by rescaling all the boundary charges by $-i$, but this spoils the Virasoro algebra.

\subsection{Tower of Higher Spins}
\label{sec25}

The calculation in Section \ref{sec24} proceeded exactly along the same lines as in \cite{Campoleoni:2010zq,Henneaux:2010xg} and should generalize to all $N$.  Indeed, once we have written the asymptotic form of the gauge field $A$ (\ref{aform}) and the gauge parameter $\xi$ (\ref{xiform}) the rest of the calculation proceeds in de Sitter exactly as in anti-de Sitter.  In particular the central charge of the ASG is related to the Chern-Simons level by the same relation, $c =6\kappa$, for any $N$.

The true case of interest is not really a bulk $SL(N,C)$ Chern-Simons theory, but Vasiliev's theory, which is based on the higher spin algebra $hs(\mu)$.  The $hs(\mu)$ algebra is infinite dimensional from the beginning (unless $\mu$ is an integer $n$, in which case the algebra truncates to $sl(n)$.)  The asymptotic symmetry analysis for this case in AdS was done in \cite{Gaberdiel:2011wb,Campoleoni:2011hg}.  This is an algebraically intensive calculation, and in this paper I will not attempt to explicitly generalize it to de Sitter.  However, the structure of the calculation presented here suggests that one can follow the derivations of \cite{Gaberdiel:2011wb,Campoleoni:2011hg} with the following identifications.  The generators of the higher spin algebra $hs(\mu)$ obey the same algebra as they do in AdS (although they may satisfy different properties under complex conjugation.)  Moreover, with the AdS metric written as
\beq
\frac{ds^2}{\ell^2} = d\rho^2 + e^{2\rho} \left(-d\tau^2 + d\theta^2\right)
\eeq
one identifies the light-cone coordinates $\tau+\theta, \tau-\theta$ with $z,\bar{z}$ and $\rho$ with the de Sitter time $t$.  Finally, one maps $\ell_{AdS}$ to $i\ell_{dS}$; this makes the CS level imaginary.  Then the rest of the computation of the ASG in de Sitter space appears to be algebraically identical to the computation in AdS.

\section{Review of $\mathcal{W}_N$ Symmetry in $AdS_3/CFT_2$}
\label{sec3}

When the bulk gravity theory resides in ${\rm AdS_3}$, one can consider the duality proposal of Gaberdiel and Gopakumar \cite{Gaberdiel:2010pz}.  The de Sitter case will be a variant of their proposal, so let us recall some of the salient features in AdS.

The CFTs of interest are WZW coset models (for a general review, see \cite{DiFrancesco:1997nk}.)  The reason for considering WZW models is that the extended Virasoro symmetries can be realized in a natural way.  Before considering the coset construction, let us recall some facts about ordinary WZW models.  One assumes that the theory contains a dimension one operator $J^a$ corresponding to a symmetry current, with OPE
\beq
J^a(z) J^b(w) \simeq \frac{k\delta^{ab}}{(z-w)^2} + \frac{i f^{ab}_{\;\;\; c} J^c(w)}{z-w}
\eeq
where the $ f^{ab}_{\;\;\; c}$ are the structure constants corresponding to some Lie algebra $g$.  In WZW models the current $J^a$ is the fundamental object and the rest of the structure of the theory is constructed from its fusions.  The natural energy-momentum tensor is the Sugawara operator
\beq
T(z) = \frac{1}{2(k + h^{\vee})} (J^a J_a)(z)
\eeq
which has the correct OPE for an energy-momentum tensor with central charge
\beq
c= {\rm dim}(g)\frac{k}{k + h^{\vee}}
\eeq
Here $h^{\vee}$ is the dual Coxeter number of the Lie algebra; for $su(N)$, $h^{\vee} = N$.

The Sugawara energy-momentum tensor exists for any current algebra, and one can think of it as corresponding to a generalization of the quadratic Casimir of the underlying Lie algebra.  Generic Lie algebras have additional Casimirs, however, and so it is natural to ask what role they play in the current algebra.  This question was first addressed in \cite{Bais:1987dc,Bais:1987zk}; for a review and further references, see \cite{Bouwknegt:1992wg}.  Specifically, one can construct operators from the normal-ordered product of several $J^a$, with Lie algebra indices contracted into the Casimirs:
\beq
d_{ab\ldots c} (J^a J^b \cdots J^c)(z)
\eeq
and compute the corresponding OPEs.  

The operators consisting of a product of $n$ $J^a$'s are spin-$n$ (the $d_{ab\ldots c}$ are always totally symmetric) and so one might hope that their OPEs are those of the extended Virasoro symmetries, but this is generically not the case.  For example, in the case of the $su(3)$ current algebra, there is a natural spin-3 operator arising from the cubic Casimir 
\beq
W \sim d_{abc} (J^a J^b J^c)(z)
\label{w3def}
\eeq
and one might hope that the current algebra has $\mathcal{W}_3$ symmetry (originally presented in \cite{Zamolodchikov:1985wn}.)  However, the $WW$ OPE is polluted by additional primary fields of dimension 4 (constructed from the normal-ordered product of four $J^a$'s) and therefore there are no extended symmetries.

Fortunately, there is a relatively simple method for constructing CFTs with $\mathcal{W}_N$ symmetry -- instead of considering simple WZW models, one can consider a coset WZW model \cite{Goddard:1984vk}.  The particular model of interest is a diagonal coset, defined as follows.  Given a Lie algebra $g$, one takes two copies of its current algebra, $J^a_{(1)}$ and $J^a_{(2)}$.  The diagonal coset is the current algebra of the sum of these currents,
\beq
J^a_{{\rm diag}} = J^a_{(1)}+J^a_{(2)}
\eeq
and is usually denoted as
\beq
\frac{g_{k_1} \oplus g_{k_2}}{g_{k_1+k_2}}
\eeq
as the level of the diagonal current algebra is $k_1+k_2$. When $g=su(N)$, the central charge corresponding to the associated Sugawara operator is
\beq
c = (N^2-1) \left(\frac{k_1}{k_1 + N}+\frac{k_2}{k_2+ N}-\frac{k_1+k_2}{k_1 + k_2+N}\right).
\eeq
For most choices of $k_1$ and $k_2$, the $WW$ OPE contains the unwanted $(J)^4$ operators, but it was shown by \cite{Bais:1987zk} that in the $su(3)$ case, when one of the levels equals is equal to 1, the quartic operators decouple due to a series of intricate cancellations.  The decoupling shows that the coset
$\frac{su(3)_k \oplus su(3)_1}{su(3)_{k+1}}$ realizes $\mathcal{W}_3$ symmetry.
The cancellations persist for cosets of the form
\beq
\frac{su(N)_k \oplus su(N)_1}{su(N)_{k+1}}
\eeq
which correspondingly exhibit $\mathcal{W}_N$ symmetry.

The authors of \cite{Gaberdiel:2010pz} suggested that the CFT dual to the Vasiliev theory in ${\rm AdS_3}$ \cite{Vasiliev:1995dn} is a minimal model associated with such an $\frac{su(N)_k \oplus su(N)_1}{su(N)_{k+1}}$ coset.  In three dimensions, the relevant version of the Vasiliev higher spin theory contains one complex scalar field of mass $M^2$ and an infinite tower of massless higher spins (with one for each integer spin $s\ge 2$.)  Because the tower of spins is infinite, one must take the limit of large $N$.  However, to obtain a large central charge (and a large de Sitter radius) it is also necessary to take the limit of large $k$.  Therefore one defines the 't Hooft-like limit $k, N \rightarrow \infty$ with the ratio $\lambda$ fixed:
\beq
\lambda = \frac{N}{k+N}.
\eeq
In this limit, the central charge is
\beq
c = \frac{3\ell}{2G} = N(1-\lambda^2).
\label{cAdS}
\eeq

The minimal CFT contains operators of dimension
\beq
\Delta_{\pm} = 1\pm \lambda
\eeq
which are dual to scalars of mass
\beq
M^2\ell^2 = -(1-\lambda^2)
\eeq
so that $-1\leq M^2\ell^2 \leq 0$.  Although the masses-squared are negative, they are above the Breitenlohner-Freedman bound.  Note also that at finite $N$, there are a variety of extra states with dimensions $h \ll N$; it was argued in \cite{Gaberdiel:2011zw} that they decouple in the large $N$ limit.

\section{A de Sitter Conjecture}
\label{sec4}
In this section we will try to find a gauge/gravity duality between a higher-spin theory in ${\rm dS_3}$ and some appropriate CFT.  It seems natural to try to construct such a theory by an appropriate modification of the WZW coset of \cite{Gaberdiel:2010pz}.  The duality proposed here is a lower dimensional counterpart of \cite{Anninos:2011ui} and the spirit of the reasoning is similar.

Our candidate higher spin theory in the bulk is Vasiliev theory in three dimensions \cite{Vasiliev:1995dn}, which can be defined in de Sitter space just as well as in anti-de Sitter space.  As in AdS, the theory contains an infinite tower of higher spin states $s=2,3,\ldots$ (one field for each integer spin greater than 1.)  In addition, there are scalar fields.  For our purposes, we take the version containing two real scalar fields of equal mass (which one can think of as a single complex scalar), as in \cite{Chang:2011mz}.  The theory is defined through a set of equations of motion; it does not have a known full Lagrangian description, although one can expand the equations of motion at a given order and write an associated effective action if desired.

The reason for choosing this exotic gravity theory is its high amount of symmetry, and we would like to see if we can use this symmetry to guess a dual CFT.  Any proposal for a dual theory is subject to a number of constraints:
\begin{itemize}
\item The central charge must be large and purely imaginary to correspond to a weakly curved bulk theory.  This implies that the CFT is necessarily non-unitary.
\item The chiral algebra of the CFT must be $\mathcal{W}_N$; if we wish to engineer this through a diagonal WZW coset, one of the levels should be equal to 1.  The other level need not be an integer, or even real, as the CFT is not unitary.  To make contact with Vasiliev's theory, we should take $N\rightarrow \infty$.
\item The CFT is Euclidean and so the chiral algebra is actually complexified.  Thus, any $su(N)$ factors in the chiral algebra are better thought of as $sl(N,C)$.
\item The operator spectrum must contain spin-zero operators dual to the scalars of Vasiliev's theory.  These scalars should  have real masses (although the CFT is not unitary, the bulk theory should be unitary.)  Using the standard dS/CFT formula for the dimensions \cite{Strominger:2001pn}, we have
\beq
\Delta_{\pm} = 1\pm \sqrt{1-M^2\ell^2}.
\eeq
If we suppose\footnote{It was emphasized to me by T. Hartman that, in the AdS case \cite{Gaberdiel:2010pz},
the allowed representations of the current algebra were explicitly computed.  Here, in contrast, the nonunitary WZW coset is not of a standard type and it is not clear whether the desired representations exist.  It is conceivable that the correct representations have a different dimension formula, giving a different constraint on the possible values of the coupling $\lambda$, or that their dimensions turn out to be incompatible with identifying the allowed operators with dual scalar fields.  Clearly, it would be interesting to determine the allowed representations rigorously.  One possible approach for doing this might be to generalize the results of \cite{Mathieu:1991fz}, who gave a prescription for finding the admissible states in nonunitary fractional WZW cosets.} that the dimensions are given (as in the AdS case \cite{Gaberdiel:2010pz}) by $\Delta_{\pm} = 1\pm \lambda$, this formula implies that $\lambda = \frac{N}{k+N}$ is either purely real or purely imaginary.  At the level of the linearized equations of motion, positivity of $M^2$ is the only dynamical constraint from the bulk theory; the higher spin fields are nonpropagating and their linearized equations are already determined by the bulk symmetries.
\item $N$, which is related to the Lie algebra Casimirs, should be real.
\end{itemize}
Taken together, these constraints are quite restrictive.

The most tempting identification, given the form of the central charge in AdS (\ref{cAdS}) is to hold $\lambda$ fixed while mapping $N \rightarrow i N$ and correspondingly $k \rightarrow ik$.  The coupling parameter $\lambda$ is then real and the masses of the scalars fall into the range $0\le M^2 \ell^2 \le 1$.  This suggestion has the appealing property that the correlation functions in AdS map to dS in a natural way.  It is conceivable that this procedure is essentially correct.  Unfortunately, I do not know of any way to accomplish this identification on the CFT side (apart from formally mapping the correlation functions) if  the current algebra is an affine Lie algebra.  

However, there is an alternative limit which does satisfy all the constraints we have listed.  If we assume that the current algebra is of standard type, then the parameter $N$ is real and a real $\lambda$ is inconsistent with an imaginary central charge.  Therefore, we have to consider the other case, where $\lambda$ is purely imaginary.  For the WZW coset
\beq
\frac{sl(N)_k \oplus sl(N)_1}{sl(N)_{k+1}}\nonumber
\eeq
one should take
\beq
k= -N + \frac{i}{\gamma}
\eeq
In the limit $N \rightarrow \infty$ and $\gamma \rightarrow \infty$, the central charge then takes the form
\beq
c = i\gamma (N^3-N) + O(\gamma^0 N^3).
\label{limitc}
\eeq
This scaling limit is evidently quite different from the usual scaling in the AdS case.
%, where one takes $k$ and $N$ to infinity with the ratio $\lambda= \frac{N}{k+N}$ fixed, and $0 \le \lambda %\le 1$.  
Note that in particular, to obtain a large central charge in our de Sitter proposal it is not necessary to take $N$ to infinity (although when $N$ is finite, we do not have a candidate for the bulk theory.)  The form of $c$ in (\ref{limitc}) is the same as in the Drinfeld-Sokolov description, where one considers an $SU(N)$ WZW model at level $k_{DS}$.  In the limit $k_{DS}\rightarrow \infty$, we have $c_N(k_{DS}) \simeq -k_{DS} N(N^2-1)$ (see Appendix B of \cite{Gaberdiel:2010pz}) so presumably one should set $k_{DS} = -i \gamma$.  This relation between $k_{DS}$ and $k$ in the coset, of course, is different from the case in ${\rm AdS_3/CFT_2}$.

%With our choice for the level $k$, $\lambda = -i\gamma N$ is purely imaginary, so that
With the assumptions we have made about the allowed operators of the CFT, the imaginary $\lambda$ implies that that the mass is large in de Sitter units:
\beq
M^2\ell^2 > 1.
\eeq
Curiously, this corresponds to the case of complex operator dimensions, 
\beq
\Delta_{\pm} = 1\mp i\gamma N,
\eeq
which are ordinarily rather confusing.  Most of the studies of scalars in dS/CFT have focused on the case of real operator dimensions and $0\le M^2\ell^2 < 1$.  It would be nice to understand the case of complex dimensions in more detail.
Although the masses are infinite in de Sitter units, they are small in Planck units.  We have $\ell \simeq \frac23 \gamma N^3 G$, so 
\beq
M \simeq \frac{3}{2N^2 G}
\eeq
That is, the large $N$ limit suppresses large quantum gravity effects in the bulk.  Finally, another way of thinking about the large $\gamma,N$ limit is to take $N\rightarrow \infty$ with $\gamma/N$ held fixed.  In this limit, we have
\beq
GM^2 \ell \sim \frac{\gamma}{N}
\eeq
The gravitational interaction between two particles at rest is proportional to $GM^2$, so this condition implies that we are holding the strength of the gravitational interaction fixed (in characteristic dS units.)

The considerations presented in this section are evidently rather schematic.  It is not at all clear that the coset CFT proposed here makes any kind of sense.  Apart from the usual concerns about non-unitary theories, one should ask whether the usual construction of the coset minimal model goes through, whether the partition function converges, and so on.  Very little is known about CFTs with complex central charge \footnote{However, see \cite{Harlow:2011ny} for a recent discussion of complex central charges in the context of Liouville theory.  I thank T. Hartman for pointing out the potential relevance of this work.}.  Hopefully the considerations presented in this paper are suggestive of what types of CFTs one should think about as possible de Sitter duals.

Of course it is also important to compute three-point functions in the bulk theory.  In ${\rm AdS_3}$, the three-point functions were computed in \cite{Chang:2011mz}, and turn out to depend on the parameters of the CFT through the quantity $\lambda$.  The proposal in this paper suggests that one should continue $\lambda$ to large imaginary values (analogous to the result in \cite{Anninos:2011ui} where one mapped $N$ to $-N$ in relating AdS to dS.) In the AdS case, the three point functions were studied in \cite{Chang:2011mz}; unfortunately their analysis is done for a specific value of the scalar mass, in the ``undeformed'' Vasiliev theory; to study the correlators relevant for our case requires a computation in the ``deformed'' Vasiliev theory.  Even the two-point functions in de Sitter space are a rich subject and it seems likely that their higher-spin analogues will be interesting too (for example, it would be interesting to revisit the description of $\alpha$-vacua \cite{Mottola:1984ar,Allen:1985ux} in the context of this CFT.)  It would also be interesting to study the semiclassical partition function of the higher spin theory directly on the bulk side, generalizing the work of \cite{Castro:2011xb}.  I intend to revisit these problems in future work \cite{nextgreatpaper}.

\section*{Acknowledgments}

I have enjoyed enlightening conversations related to this work with Alejandra Castro, Keshav Dasgupta, Tom Hartman, Martin Kruczenski, Nima Lashkari, Alex Maloney, and Georgios Michalogiorgakis.  I especially thank T. Hartman for very useful correspondence, as well as critical comments on a draft version of this paper.  I would also like to thank the McGill High Energy Theory group for its warm hospitality while this work was in progress.  This work was supported in part by the DOE under grant DE-FG02-91ER40681.

%\appendix

\bibliography{higherspin}\bibliographystyle{utphys}

\end{document}